\documentclass[conference,twocolumn,a4paper]{IEEEtran}
\usepackage{cite}
\usepackage{amsmath,amssymb,amsfonts}
\usepackage{graphicx}
\usepackage{tikz}
\usetikzlibrary{IBG}
\usepackage{pgfplots}
\usetikzlibrary{plotmarks}
\usetikzlibrary{arrows}
\usetikzlibrary{spy,backgrounds}
\usepackage{tikz-3dplot}
\usetikzlibrary{positioning,chains,fit,shapes,calc}
\usetikzlibrary{pgfplots.groupplots}
\usepackage{forloop}
\usepackage{tkz-tab} 
\usepackage{diagbox}
\usepackage{subcaption}
\pgfplotsset{compat=newest}

\newcommand*\circled[1]{\tikz[baseline=(char.base)]{
            \node[shape=circle,draw,inner sep=0pt] (char) {#1};}}

\definecolor{DarkBlue}{rgb}{0, 0.2706, 0.541}
\definecolor{DarkOrange}{rgb}{0.92, 0.43, 0}
\definecolor{DarkGreen}{rgb}{0.098, 0.4784, 0.5176}
\definecolor{DarkRed}{rgb}{0.7529, 0, 0}

\begin{document}
\bstctlcite{IEEEexample:BSTcontrol} 
%

	\title{Decoding of Non-Binary LDPC Codes Using the Information Bottleneck Method}
	\author {
		\makebox[.5\linewidth]{Maximilian Stark, Gerhard Bauch}\\ \textit{Hamburg University of Technology} \\	\textit{Institute of Communications}\\			21073 Hamburg, Germany\\			\{maximilian.stark,bauch\}@tuhh.de\\
		\and \makebox[.5\linewidth]{Jan Lewandowsky, Souradip Saha}\\\textit{Fraunhofer Institute for Communication,}\\	\textit{Information Processing and Ergonomics (FKIE)}\\		53343 Wachtberg, Germany\\				\{jan.lewandowsky,souradip.saha\}@fkie.fraunhofer.de%
		\thanks{This work has been accepted for publication at the IEEE International Conference on Communications (ICC'19) in Shanghai.}
	}
	
	\maketitle
	
	\begin{abstract}
Recently, a novel lookup table based decoding method for binary low-density parity-check codes has attracted considerable attention. In this approach, mutual-information-maximizing lookup tables replace the conventional operations of the variable nodes and the check nodes in message passing decoding. Moreover, the exchanged messages are represented by integers with very small bit width. A machine learning framework termed the information bottleneck method is used to design the corresponding lookup tables. In this paper, we extend this decoding principle from \textit{binary} to \textit{non-binary} codes. This is not a straightforward extension but requires a more sophisticated lookup table design to cope with the arithmetic in higher order Galois fields. Provided bit error rate simulations show that our proposed scheme outperforms the log-max decoding algorithm and operates close to sum-product decoding.
\end{abstract}
	
	
	\section{Introduction}  
	Shortly after their rediscovery by MacKay \cite{MacKay.1998}, binary low-density parity-check (LDPC) codes have been generalized for non-binary symbol alphabets over higher order Galois fields with field order $q$. However, the decoding of these codes using sum-product decoding is computationally much more expensive than the decoding of their binary counterparts.
	The main computational bottleneck for higher order Galois field LDPC codes is the required convolution of probability distributions at the check nodes. Moreover, the number of bits required to represent the processed probability vectors in hardware is large.
	Approaches to reduce the implementation complexity of the check node operation range from application of the fast convolution using the fast Walsh-Hadamard transform (FHT) to 
	log-domain decoding with an approximated check node operation \cite{Savin.2008,Declercq.2007,Declercq.2005,Wymeersch.2004}. 
	Despite these very important works, the development of efficient decoding methods for non-binary LDPC codes continues to be an interesting subject of current research for practical purposes as non-binary LDPC codes have better error correction properties for short block lengths than binary LDPC codes. The latter unfold their capacity approaching behavior only for very large codeword lengths \cite{Chung.2001}.
	Therefore, especially in 5G related scenarios such as massive-machine-type communications and ultra-reliable low latency communication (uRLLC), non-binary LDPC codes could be promising candidates, if decoders with affordable complexity were available.
	
	Recently, several authors considered a novel approach for decoding of \textit{binary} LDPC codes 
	\cite{Lewandowsky.2015,Romero.2016,Meidlinger.2015,Lewandowsky.2018,Stark.2018c,applsci-355896}. In these works, a principle was applied which is fundamentally different from state-of-the-art signal processing approaches. Instead of implementing the sum-product algorithm to decode an LDPC code, mutual-information-maximizing lookup tables were used to replace all conventional signal processing steps in an LDPC decoder. These lookup tables process only quantization indices which can be stored using just a few bits in hardware. Moreover, all expensive operations were replaced by simple lookup operations in the designed mutual-information-maximizing lookup tables. In \cite{Ghanaatian.2017} and \cite{Lewandowsky.2018b} it was shown that this approach is, in fact, beneficial in comparison to state-of-the-art LDPC decoders in practical decoder implementations. 
	The applied mutual-information-maximizing lookup tables can be constructed using the information bottleneck method \cite{Tishby.1999}. 
	
	The existing works \cite{Lewandowsky.2015,Romero.2016,Meidlinger.2015,Lewandowsky.2018,Stark.2018c,applsci-355896} only describe the decoding of \textit{binary} LDPC codes with the proposed method. In this paper, our aim is to extend the fundamental principle also to \textit{non-binary} LDPC codes. This extension is not straightforward as it requires sophisticated lookup table design approaches. 
	From the results obtained, we observe that the proposed algorithm performs very close to to the sum-product algorithm. 
	
	The paper contains the following main contributions:
	\begin{itemize}
		\item We devise relevant-information-preserving variable and check node operations using the information bottleneck method resulting in a novel decoder for non-binary LDPC codes. 
		\item In the resulting decoder, all arithmetic operations are replaced by simple lookups.		
		\item This novel decoder can be applied for arbitrary regular non-binary LDPC codes.
		\item Inherently, we devise a discrete density evolution scheme for non-binary LDPC codes which can be used to study the performance of non-binary code ensembles under the considered lookup table decoding.
		\item Despite all operations being simple lookup operations and all messages being passed during decoding are represented with a few bits, our proposed decoder shows only 0.15 dB performance degradation over $E_b/N_0$ compared to double-precision sum-product decoding and outperforms double-precision log-max decoding by 0.4 dB for an exemplary code over GF$(4)$.
	\end{itemize}

	The rest of this paper is structured as follows. The following section introduces the prerequisites on non-binary LDPC codes,
	the considered transmission system and the mutual-information-maximizing lookup table design with the information bottleneck method. Section \ref{sec:main} compares the conventional sum-product algorithm used for the decoding of non-binary LDPC codes with the proposed lookup table based approach in detail. In Section \ref{sec:results} we investigate the performance of the proposed approach for an exemplary code. Finally, we conclude the paper in Section \ref{sec:conclusion}.
	
	\subsubsection*{Notation}
	The realizations $y \in \mathcal{Y}$ from the event space $\mathcal{Y}$ of a discrete random variable $Y$ occur with probability $\Pr(Y=y)$ and $p(y)$ is the corresponding probability distribution. The cardinality or alphabet size of a random variable is denoted by $|\mathcal{Y}|$. Joint distributions and conditional distributions are denoted $p(x,y)$ and $p(x|y)$, $\delta(x)$ denotes the Kronecker delta function.

	\section{Prerequisites}
	\label{seq:prerequisites}
	
	This section briefly reviews non-binary LDPC codes and the information bottleneck method. Then, decoding of binary LDPC codes using the information bottleneck is summarized.
		
	\subsection{Non-binary LDPC Codes}
	LDPC codes are typically defined using a sparse parity-check matrix $\mathbf{H}$ with dimension $N_\text{c}\times N_\text{v}$ such that a parity-check equation for a codeword $\mathbf{c}$ fulfills
	$
	 \mathbf{H}\cdot \mathbf{c}=\mathbf{0}.
	$
    Each row of $\mathbf{H}$ represents a parity-check equation. Such an equation has the form
    \begin{equation}
     \sum\limits_{k=0}^{d_\text{c}-1} \underbrace{h_kc_k}_{c'_k} = 0,
     \label{eq:parity_check_eq}
    \end{equation}
    where $h_k$ correspond to the non-zero entries of the respective row, $c_k$ are the corresponding codeword symbols and $d_\text{c}$ denotes the check node degree.
    The arithmetic that has to be applied in (\ref{eq:parity_check_eq}) depends on the field order of the considered Galois field. For binary codes, all $h_k=1$, all $c_k\in\text{GF}(2)=\{0,1\}$ and the sum is a modulo $2$ sum. In contrast, in the non-binary case all $h_k$ and $c_k$ and hence their products $c_k'=h_kc_k$ too are field elements from GF($q$).
    Therefore, the arithmetical rules for multiplication and addition for the respective finite field have to be taken into account. We consider extension fields $\text{GF}(2^m)=\{0,1,\alpha,\alpha^2,\ldots,\alpha^{2^m-2}\},$ where $\alpha$ is the so-called primitive element of the field. Such a field is generated by a primitive polynomial. 
    The primitive polynomial can be used to derive multiplication and addition rules for two given elements $c_i,c_j\in\text{GF}(2^m)$. These rules are exemplarily shown for $\text{GF}(2^2)$ in Table \ref{tab:rules}.
    For their exact derivation and more details on the corresponding field theory, we refer the reader to \cite{Carrasco.2008}.
    
	\begin{table}
	\centering
	\caption{Arithmetic in GF$(2^2)$}
	\label{tab:rules}
	 \begin{minipage}{0.48\columnwidth}
	 \centering
	 \caption*{Addition $c_i+c_j$}
	 \setlength{\tabcolsep}{5pt} 
     \renewcommand{\arraystretch}{1.2} 
	 \begin{tabular}{c||c|c|c|c}	 
	 \backslashbox[10mm]{$c_i$}{$c_j$}&$0$ & $1$ & $\alpha$	& $\alpha^2$	\\
	 \hline\hline
	 $0$ &$0$ & $1$ & $\alpha$	& $\alpha^2$\\
	 \hline
	 $1$ &$1$ & $0$ & $\alpha^2$	& $\alpha$\\
	 \hline
	 $\alpha$ &$\alpha$ & $\alpha^2$ & $0$	& $1$\\
	 \hline
	 $\alpha^2$ &$\alpha^2$ & $\alpha$ & $1$	& $0$
	  \end{tabular}	 	  
	 \end{minipage}
    \begin{minipage}{0.48\columnwidth}    
    \centering    
    \caption*{Multiplication $c_ic_j$}
    \setlength{\tabcolsep}{5pt} 
     \renewcommand{\arraystretch}{1.2} 
	 \begin{tabular}{c||c|c|c|c}
	 \backslashbox[10mm]{$c_i$}{$c_j$}&$0$ & $1$ & $\alpha$	& $\alpha^2$	\\
	 \hline\hline
	 $0$ &$0$ & $0$ & $0$	& $0$\\
	 \hline
	 $1$ &$0$& $1$ & $\alpha$	& $\alpha^2$\\
	 \hline
	 $\alpha$ &$0$ & $\alpha$ & $\alpha^2$	& $1$\\
	 \hline
	 $\alpha^2$ &$0$ & $\alpha^2$ & $1$	& $\alpha$ 
	  \end{tabular}	 
	 \end{minipage}
	 \vskip-3pt  
    \end{table}

\subsection{The Information Bottleneck Method and Relevant-Information-Preserving Signal Processing}
	The information bottleneck method \cite{Tishby.1999} is a mutual-information-maximizing clustering framework 
	from machine learning. As depicted in Figure \ref{fig:IB} 
	\begin{figure}
        \begin{subfigure}{0.45\columnwidth}
        \centering  
        \begin{tikzpicture}
 
 \node (X) {$X$};
 \node[right = 0.7cm of X] (Y) {$Y$};
 \node[right = 0.7cm of Y] (T) {$T$};

 \draw[->,>=stealth] (X) --  node[] {} (Y);
 \draw[->,>=stealth] (Y) --  node[label= below:$\min\limits_{p(t|y)} \text{I}(Y;T)$] {} (T);
 
 \draw[<->,>=stealth] (X)  to [bend left=30] node[pos=0.5,above] {$\max\limits_{p(t|y)} \text{I}(X;T)$}  (T);
 
\end{tikzpicture}
        \caption{}
        \label{fig:IB} 
        \end{subfigure}
        \begin{subfigure}{0.45\columnwidth}
        \centering 
        \begin{tikzpicture}
 
\matrix [row sep=2mm,column sep=5mm] {
	\node[latent]	(y0) {$y_0$}; \\[-5mm]
	&  \IBGnode {IBN01} {$x$} {} {}; & \node[latent]	(t1) {$t$}; \\[-5mm]
	\node[latent]	(y2) {$y_1$}; \\[-1.5mm]
};


\graph [use existing nodes] {
	
	y0 -- IBN01 -- t1 ;
	y2 -- IBN01 -- t1 ;
	};

\end{tikzpicture}
        \caption{}
        \label{fig:LUT}  
        \end{subfigure}   
        \vskip-3pt                
        \caption{(a) Illustration of the information bottleneck principle, (b) Exemplary information bottleneck graph.}
        \vskip-15pt                
    \end{figure}
	it considers a Markov chain $X\rightarrow Y \rightarrow T$ of three random variables.
	$X$ is termed the relevant variable, $Y$ is termed the observation and $T$ is a compressed representation of $Y$. The compression is described by the conditional distribution $p(t|y)$. This compression mapping is designed such that the mutual information I$(X;T)$ is maximized while at the same time the mutual information I$(Y;T)$ is minimized. If the 
	mapping $p(t|y)$ uniquely assigns a $t$ to each $y$ with probability $1$, this mapping can be implemented in a lookup table such that $t=f(y)$. Algorithms to find suitable compression mappings are described in \cite{Slonim.2002, Hassanpour.2017,SL18}. These algorithms require the joint distribution $p(x,y)$ and the desired cardinality $|\mathcal{T}|$ of the compression variable $T$ as inputs. As a by-product, an information bottleneck algorithm delivers the joint distribution $p(x,t)=p(x|t)p(t)$.
	
	Preliminary works have shown that the information bottleneck design principle can be applied to build signal processing blocks which implement mutual-information-maximizing lookup tables. Figure \ref{fig:LUT} shows such a mutual-information-maximizing lookup table. This figure uses the information bottleneck graph notation introduced in \cite{Lewandowsky.2016}. The inputs $(y_0,y_1)$ of the shown lookup table are compressed by the mutual-information-maximizing lookup table such that the output $t$ is highly informative about the relevant variable $X$. 
	The processed inputs $(y_0,y_1)$ and the output $t$ of the system are considered to be quantization indices from the set $\mathcal{T} = \{0,1,\ldots,|\mathcal{T}|-1\}$. In contrast, state-of-the-art signal processing algorithms process quantized samples with a certain precision. However, in the information bottleneck approach this is not required as the mutual information of the involved variables does not depend on the representation values of the quantized signal, but only on their joint probability distributions.
\subsection{Information Bottleneck Decoding of Binary LDPC Codes}
In prior works on binary LDPC codes, the relevant variable $X$ for a check node is the modulo $2$ sum of the bits connected to the check node. Thus, the mutual-information-maximizing lookup table serves as an integer-based replacement for the well-known box-plus operation for log-likelihood ratios. The approach in prior works was to construct lookup tables for each node type and every iteration using a framework which pairs a density evolution technique with an information bottleneck algorithm \cite{Lewandowsky.2015,Romero.2016,Meidlinger.2015,Lewandowsky.2018,Stark.2018c,applsci-355896}. This complex lookup table construction step is performed offline. The lookup tables are pre-generated for a fixed design-$E_b/N_0$, but used for all $E_b/N_0$ in practice. Once constructed, all decoding operations become lookup operations in the pre-generated tables. This approach achieves considerable gains in decoding throughput \cite{Ghanaatian.2017} and performance extremely close to double-precision sum-product decoding for binary codes.
In the following section, we present all the required steps to generalize the construction framework \cite{Lewandowsky.2018} from binary to non-binary LDPC codes. This generalization is not straightforward and the challenges are versatile. The main reason is the much more sophisticated arithmetic in higher order Galois fields.

\section{Decoding Non-Binary LDPC codes using the Information Bottleneck Method}
\label{sec:main}
In this section, we describe how a lookup table based decoder for non-binary LDPC codes is built. In each step we start from the conventional sum-product decoding and compare it with the lookup table approach. We first describe the transmission scheme and the channel output quantizer. Then, we explain how check and variable node operations in the sum-product algorithm can be replaced by lookup tables.

\subsection{Transmission Scheme and Channel Output Quantization}
\label{sec:qtzr}
We consider a non-binary LDPC encoded transmission over a quantized output, symmetric additive white Gaussian noise (AWGN) channel with binary phase shift keying modulation (BPSK). In the applied scheme, $m$ BPSK symbols are transmitted for each codeword symbol $c_k$. At the receiver, the received signal is first quantized. The quantizer delivers $m$ outputs $\mathbf{y}_k=[y_{k,0},y_{k,1},\ldots,y_{k,m-1}]^\text{T}$ for each codeword symbol $c_k$. The bit width of the applied quantizer is 
denoted $w$, such that the outputs $y_{k,j}$ are from alphabet $\{0,1,\ldots,2^w-1\}$. 

The first step in conventional sum-product decoding of non-binary LDPC codes is the calculation of the 
symbol probabilities
\begin{equation}
 p(c_k|\mathbf{y}_k)=\frac{p(c_k)}{p(\mathbf{y}_k)}\prod\limits_{j=0}^{m-1}p(y_{k,j}|b_{k,j}),
 \label{eq:combi}
\end{equation}
where $b_{k,j}$ denote the bits in the binary representation of $c_k$.
For each symbol, this corresponds to a probability vector which is used as channel knowledge for sum-product decoding.

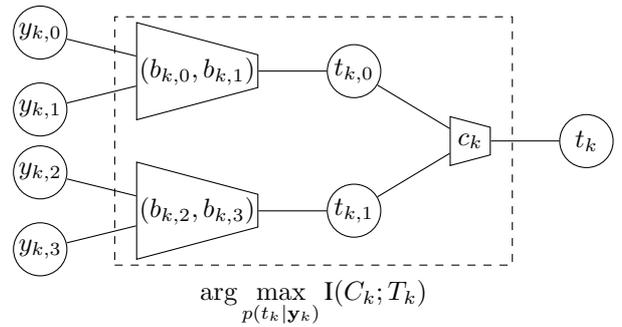
\begin{figure}
 \centering 
 \begin{tikzpicture}[hv path/.style={to path={-| (\tikztotarget)}},
vh path/.style={to path={|- (\tikztotarget)}}]

\matrix [row sep=2mm,column sep=9mm] {
	\node[latent]	(y1) {$y_{k,0}$};& & &; \\[-7mm]
	&  \IBGnode {IBN11} {$(b_{k,0},b_{k,1})$} {} {}; &  \node[latent] (t11) {$t_{k,0}$}; \\[-7mm]
	\node[latent]	(y2) {$y_{k,1}$};& & &; \\[-5mm]
	& & &; \IBGnode {IBN_21} {$c_k$} {} {}; & \node[latent] (t21) {$t_k$};\\[-5mm]
	\node[latent]	(y3) {$y_{k,2}$};& & &; \\[-7mm]
	&; \IBGnode {IBN_12} {$(b_{k,2},b_{k,3})$} {} {}; & \node[latent] (t12) {$t_{k,1}$}; \\[-7mm]
	\node[latent]	(y4) {$y_{k,3}$};& & &  \\
};

\node[draw, rectangle,dashed, fit=(IBN_12)(IBN_21)(IBN11), inner sep=8pt] (plate) {};

\node[below= 2pt of plate.south] () {$\arg\max\limits_{p(t_k|\mathbf{y}_k)} \text{I}(C_k;T_k)$};


\graph [use existing nodes] {
	y1 --IBN11 --t11 ;
	y2 --IBN11;
	y3 --IBN_12 --t12 ;
	y4 --IBN_12;
	t12 --IBN_21 --t21 ;
	t11 --IBN_21;
};

\end{tikzpicture}
 \vskip -8pt   
 \caption{Information bottleneck graph of lookup table $p(t_k|\mathbf{y}_k)$. }
 \label{fig:IBG_combi} 
 \vskip -6pt
\end{figure}

In contrast, the proposed information bottleneck decoder does not use any probability vector, but processes a single quantization index $t_k\in\{0,1,\ldots,|\mathcal{T}_\text{chan}|-1\}$ instead. Intuitively, this quantization index should be highly informative about $c_k$. Such an index $t_k$ can be obtained from $\mathbf{y}_k$ using a mutual-information-maximizing lookup table $p(t_k|\mathbf{y}_k)$ which is constructed with the information bottleneck method. The required joint distribution $p(c_k,\mathbf{y}_k)$ to construct the table follows directly from (\ref{eq:combi}). 
As a by-product we obtain $p(c_k,t_k)$ which will be used for the construction of subsequent lookup tables.
The size of the lookup table can be reduced by using a decomposition into two-input lookup tables as exemplified in Figure \ref{fig:IBG_combi} for $m=4$ inputs. 
\subsection{Non-Binary Check Node Operations in Sum-Product Decoding}
In sum-product decoding of non-binary LDPC codes, the symbol probabilities (\ref{eq:combi}) are passed to the check nodes. Each check node performs three tasks according to its parity-check equation (\ref{eq:parity_check_eq}).

\subsubsection*{1) Multiplication by Edge Weights $c_k'=h_kc_k$}
First, the incoming probability vectors for the incoming symbols $c_k$ are transformed into the probability vectors for the products $c_k'$ incorporating the appropriate edge weight $h_k$. According to the multiplication rules described in Section \ref{seq:prerequisites}, this corresponds to a cyclic shift of the last $2^m-1$ entries in the probability vectors \cite{Carrasco.2008}. 

\subsubsection*{2) Summation}
Once all $p(c_k')$ are obtained, the check node computes the convolution of $d_\text{c}-1$ probability vectors $p(c_k')$ to account for the summation of the involved $c_k'$ in  $c_{j}'=\sum_{k\neq j}c_k'$ which follows from (\ref{eq:parity_check_eq}). This convolution is usually implemented as a fast convolution using FHT resulting in the complexity $\mathcal{O}(d_\text{c}2^m\log_2 2^m)$. 
\subsubsection*{3) Multiplication by Inverse Edge Weights $c_j=h_j^{-1}c_j'$}
In the last step of the check node update, the outgoing message which is passed to a connected variable node is again found by a cyclic shift of the last $2^m-1$ entries of $p(c_j')$ according to the inverse edge weight $h_j^{-1}$.

\subsection{Non-Binary Check Node Operations from the Information Bottleneck Method}
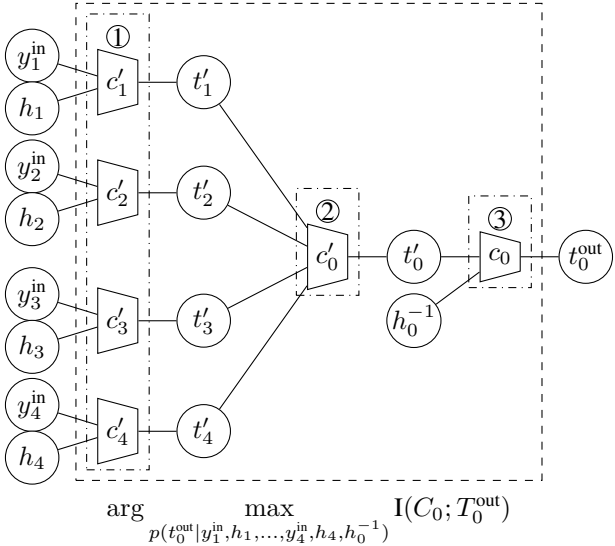
\begin{figure}
 \centering 
 \begin{tikzpicture}[hv path/.style={to path={-| (\tikztotarget)}},
vh path/.style={to path={|- (\tikztotarget)}}]

\matrix [row sep=2mm,column sep=5mm] {
	\node[latent]	(y0) {$y_1^\text{in}$}; \\[-6.5mm]
	&  \IBGnode {IBN01} {$c'_{1}$} {} {}; & \node[latent]	(tprime0) {$t'_1$};& & &; \\[-6.5mm]
	\node[latent]	(h0) {$h_1$};\\[-1.5mm]
	\node[latent]	(y1) {$y_2^\text{in}$};\\[-6.5mm]
	& \IBGnode {IBN02} {$c'_{2}$} {} {};  & \node[latent]	(tprime1) {$t'_2$};& & &; \\[-6.5mm]
	\node[latent]	(h1) {$h_2$};\\[-5mm]
	& & & &; \IBGnode {IBN_21} {$c_0'$} {} {}; & \node[latent] (t21) {$t_0'$};&  \IBGnode {IBN_31} {$c_{0}$} {} {};& \node[latent] (t31) {$t_0^\text{out}$};& \\[-5mm]
	\node[latent]	(y2) {$y_3^\text{in}$};& & & & & \\[-6.5mm]
	& \IBGnode {IBN03} {$c'_{3}$} {} {}; & \node[latent]	(tprime2) {$t'_3$};& & &\node[latent]	(hinv) {$h_0^{-1}$}; \\[-6.5mm]
	\node[latent]	(h2) {$h_3$};\\[-1.5mm]
	\node[latent]	(y3) {$y_4^\text{in}$}; \\[-6.5mm]
	& \IBGnode {IBN04} {$c'_{4}$} {} {};  & \node[latent]	(tprime3) {$t'_4$};& & & & \\[-6.5mm]
	\node[latent]	(h3) {$h_4$};\\[-5mm]
};


\graph [use existing nodes] {
	
	y0 -- IBN01 -- tprime0 -- IBN_21 --t21 --IBN_31-- t31 ;
	h0 -- IBN01;
	y1 -- IBN02 -- tprime1 -- IBN_21;
	h1 -- IBN02;
	y2 -- IBN03 -- tprime2 -- IBN_21;
	h2 -- IBN03;
	y3 -- IBN04 -- tprime3 -- IBN_21;
	h3 -- IBN04;
	hinv -- IBN_31;
	};

\node[above= 2pt of IBN01.west,inner sep=0pt] (circ1) {\circled{1}};
\node[draw, rectangle,dashdotted, fit=(IBN01)(IBN04)(circ1), inner sep=4pt] (plate) {};

\node[above= 2pt of IBN_21.west,inner sep=0pt] (circ2) {\circled{2}};
\node[draw, rectangle,dashdotted, fit=(IBN_21)(circ2), inner sep=4pt] (plate2) {};

\node[above= 2pt of IBN_31.west,inner sep=0pt] (circ3) {\circled{3}};
\node[draw, rectangle,dashdotted, fit=(IBN_31)(circ3), inner sep=4pt] (plate3) {};
	
\node[draw, rectangle,dashed, fit=(IBN01)(IBN04)(IBN_31)(circ1), inner sep=8pt] (plate4) {};
\node[below= 2pt of plate4.south] () {$\arg\max\limits_{p(t_0^\text{out}|y_1^\text{in}, h_1, \ldots, y_{4}^\text{in},h_4,h_0^{-1})} \text{I}(C_0;T_0^\text{out})$};


\end{tikzpicture}
 \vskip -2pt               
 \caption{Information bottleneck graph of lookup table $p(t_0^\text{out}|y_1^\text{in}, h_1, \ldots, y_4^\text{in},h_4,h_0^{-1})$ for $d_\text{c} =5$.}
 \label{fig:check} 
  \vskip -8pt  
\end{figure}

Here, we propose to replace all of the aforementioned operations with mutual-information-maximizing lookup tables. 
The entire workflow of the check node design with the information bottleneck method is exemplified in Figure \ref{fig:check} for a degree $d_\text{c}=5$ check node. 
This check node processes $d_\text{c}-1=4$ incoming quantization indices  $y_k^\text{in}$ to determine one outgoing quantization index $t_0^\text{out}$ which is passed back to the variable node replacing 
the probability vector $p(c_0)$ in the sum-product algorithm. 
Please note that the message $y_0^\text{in}$ for $c_0$ is excluded 
since extrinsic information on $c_0$ shall be generated. Message generation has to be carried out using an equivalent structure for all other $c_j$.

We provide a step-by-step derivation of the joint distributions required as inputs for the information bottleneck algorithms to generate the respective mutual-information-maximizing lookup tables.

\subsubsection*{1) Multiplication by Edge Weights $c_k'=h_kc_k$} 
In Figure \ref{fig:check} the multiplication equivalent lookup table is depicted in the box labeled \circled{1}. Obviously, since all incoming quantization indices $y_k^\text{in}$ are just unsigned integers, no shift of any probability vector is possible. However, this is not required since we are only interested in preserving the information on the relevant random variable $C'_k$ given the input tuple $(h_k,y_k^\text{in})$. Therefore, we need to determine the joint distribution $p(c_k',h_k,y_k^\text{in})$ to design a mutual-information-maximizing lookup table with the information bottleneck method. According to the general chain rule of probabilities and given the independence of $y_k^\text{in}$ and $h_k$, 
\begin{equation}
 p(c_k',h_k,y_k^\text{in})=\sum\limits_{c_k\in\text{GF}(2^m)}p(c_k'|h_k,c_k)p(c_k,y_k^\text{in})p(h_k).
 \label{eq:joint_prod}
\end{equation}
In (\ref{eq:joint_prod}), $p(c_k'|h_k,c_k)$ reflects the multiplication arithmetic $c_k'=h_kc_k$ in GF$(2^m)$. Mathematically,
$p(c_k'|h_k,c_k)=\delta(c_k'+h_kc_k)$, i.e., it is $1$ if $c_k'=h_kc_k$ and $0$ otherwise.
In the first decoding iteration, $p(c_k,y_k^\text{in})$ is given by $p(c_k,t_k)$ with $t_k=y_k^\text{in}$ since all incoming $y_k^\text{in}$ are obtained directly from the quantizer (cf. Section \ref{sec:qtzr}).
Feeding the joint distribution (\ref{eq:joint_prod}) to an information bottleneck algorithm with output cardinality $|\mathcal{T}_\text{mult}|$ delivers the lookup table $p(t_k'|h_k,y_k^\text{in})$, where $t_k'\in\{0,1,\ldots,|\mathcal{T}_\text{mult}|-1\}$ and I$(C'_K;T_K')\rightarrow\max$ for the given cardinality $|\mathcal{T}_\text{mult}|$.
\subsubsection*{2) Summation}
To account for the summation $c_0'=\sum_{k\neq 0}c_k'$, in Figure \ref{fig:check}, the convolution equivalent lookup table is depicted in the box labeled \circled{2}. 
Again since only unsigned integers $t_k'$ are processed instead of probability vectors, 
a new $t_0'$ given $(t_1',t_2'\ldots,t_{d_\text{c}-1}')$ has to be generated which is highly informative about $c_0'=\sum_{k\neq 0}c_k'$.
Therefore, we need the joint distribution $p(c_0',t_1',t_2'\ldots,t_{d_\text{c}-1}')$.
Similarly as in (\ref{eq:joint_prod}) one finds
\begin{multline}
 p(c_0',t_1',t_2',\ldots,t_{d_\text{c}-1}')=\\
 \sum\limits_{c_1',c_2',\ldots,c_{\text{d}_c-1}'} p(c_0'|c_1',c_2',\ldots,c_{d_\text{c}-1}')\prod\limits_{k=1}^{d_\text{c}-1}p(t_k',c_k'). 
 \label{eq:joint_conv}
\end{multline}  
In (\ref{eq:joint_conv}), $p(c_0'|c_1',c_2',\ldots,c_{d_\text{c}-1}')$ reflects the sum arithmetic $c_0'=\sum_{k\neq 0}c_k'$ in GF$(2^m)$. Mathematically,
$p(c_0'|c_1',c_2',\ldots,c_{d_\text{c}-1}')=\delta(c_0'+\sum_{k\neq 0}c_k')$. 
Feeding the joint distribution (\ref{eq:joint_conv}) to an information bottleneck algorithm with output cardinality $|\mathcal{T}_\text{conv}|$ delivers a lookup table $p(t'_0|t_1',t_2',\ldots,t_{\text{d}_c-1}')$, where $t'_0\in\{0,1,\ldots,|\mathcal{T}_\text{conv}|-1\}$ and I$(C'_0;T'_0)\rightarrow\max$ for the given cardinality $|\mathcal{T}_\text{conv}|$.

Similar as in Section \ref{sec:qtzr}, we note that a two-input decomposition of lookup tables can be applied to reduce the size of the lookup table $p(t'_0|t_1',t_2',\ldots,t_{\text{d}_c-1}')$.

\subsubsection*{3) Multiplication by Inverse Edge Weights $c_0=h_0^{-1}c_0'$}
The multiplication equivalent by the inverse edge label $h_0^{-1}$ is also implemented as a mutual-information-maximizing lookup table $p(t_0^\text{out}|h_0^{-1},t_0')$ and depicted in the box labeled \circled{3} in Figure \ref{fig:check}. The joint distribution $p(c_0,h_0^{-1},t_0')$ for designing the
involved lookup table can be obtained equivalently as explained for the multiplication equivalent by $h_0$ using (\ref{eq:joint_prod}).
The final output $t_0^\text{out}\in\{0,1,\ldots,|\mathcal{T}_\text{prod}|-1\}$ is passed to a connected variable node.

\subsection{Non-Binary Variable Node Operations in Sum-Product Decoding}
In sum-product decoding of non-binary LDPC codes, each variable node receives $d_\text{v}$ probability vectors from its connected check nodes, where $d_\text{v}$ is the degree of the variable node.
To generate extrinsic information which is passed back to the check nodes during decoding, $d_\text{v}-1$ messages from the check nodes and the channel message (\ref{eq:combi}) are multiplied. This results from the equality constraint of a variable node, i.e., all incoming messages are probability vectors for the same codeword symbol. 

\subsection{Non-Binary Variable Node Operations from the Information Bottleneck Method}
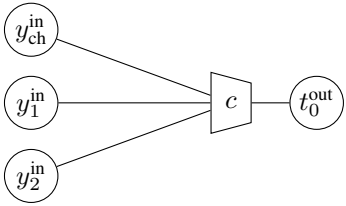
\begin{figure} 
 \centering 
 \begin{tikzpicture}[hv path/.style={to path={-| (\tikztotarget)}},
vh path/.style={to path={|- (\tikztotarget)}}]

\matrix [row sep=2mm,column sep=5mm] {
	\node[latent]	(y0) {$y_\text{ch}^\text{in}$};& & & &; \\
	\node[latent]	(y1) {$y_1^\text{in}$};& & & &; \IBGnode {IBN_21} {$c$} {} {}; & \node[latent] (t21) {$t_0^\text{out}$};& \\
	\node[latent] (y2) {$y_2^\text{in}$};& & & &; \\
};


\graph [use existing nodes] {
	y0 -- IBN_21 --t21 ;
	y1 -- IBN_21;
	y2 -- IBN_21;
	};



\end{tikzpicture} 
 \vskip -2pt    
 \caption{Information bottleneck graph of lookup table $p(t_0^\text{out}|y_\text{ch}^\text{in},y_1^\text{in}, y_2^\text{in}, \ldots, y_{d_\text{v}-1}^\text{in})$ for $d_\text{v}=3$.}   
 \label{fig:var}  
 \vskip -8pt
\end{figure}

In the following, we consider an arbitrary node which belongs to a codeword symbol $c$. 
Here, we propose to replace the described variable node operation with a mutual-information-maximizing lookup table. This lookup table is depicted in Figure \ref{fig:var} and it processes $d_\text{v}-1$ incoming quantization indices $y_k^\text{in}$ received from the check nodes and a channel index $y_\text{ch}^\text{in}$ from the channel output quantizer to determine one outgoing quantization index $t_0^\text{out}$ which is passed back to a check node. 
Please note that the message $y_0^\text{in}$ is excluded at the input on the left since extrinsic information shall be generated. Message generation has to be carried out with the same structure for all other connected edges.
The joint input distribution to design the depicted lookup table in Figure \ref{fig:var} is given by
\begin{equation}
 p(c,y_\text{ch}^\text{in},y_1^\text{in},y_2^\text{in},\ldots,y_{d_\text{v}-1}^\text{in})=p(c)p(y_\text{ch}^\text{in}|c)\prod\limits_{l=1}^{d_\text{v}-1}p(y_l^\text{in}|c).
 \label{eq:joint_var}
\end{equation}
This joint distribution reflects the aforementioned equality constraint of the variable node.
Feeding the joint distribution (\ref{eq:joint_var}) to an information bottleneck algorithm with output cardinality $|\mathcal{T}_\text{var}|$ delivers a lookup table $p(t_0^\text{out}|y_\text{ch}^\text{in},y_1^\text{in},y_2^\text{in},\ldots,y_{d_\text{v}-1}^\text{in})$, where $t_0^\text{out}\in\{0,1,\ldots,|\mathcal{T}_\text{var}|-1\}$ and I$(C;T_0^\text{out})\rightarrow\max$. The unsigned integer $t_0^\text{out}$ is passed back to the connected check node on the target edge in the next decoding iteration.

Finally, we note that a two-input decomposition of lookup tables can be applied to reduce the size of the variable node lookup table.
\subsection{Discrete Density Evolution for Non-Binary Codes and Fixed Lookup Tables}
It is important that the distributions of the exchanged messages evolve over the iterations. To cope with this evolution it is, therefore, appropriate to design updated lookup tables for each decoding iteration using the appropriate distributions. These joint distributions correspond to the by-products $p(x,t)$ of the applied information bottleneck algorithm.
By using these output distributions as inputs of the next applied information bottleneck to construct lookup tables, we inherently track the evolution of these joint input distributions. This is completely analogous to the discrete density evolution scheme for binary LDPC codes described in \cite{Romero.2016,Meidlinger.2015,Lewandowsky.2018,Stark.2018c}. As an interesting consequence, the decoding performance for a considered regular ensemble under the proposed lookup table based decoding scheme can be investigated. We note that performing efficient density evolution for non-binary LDPC codes is an open problem which is inherently tackled by the proposed lookup table construction scheme. However, since this is not the main topic of this paper, we defer further investigation of this interesting finding to a subsequent work.

Finally, we propose to construct all involved lookup tables just once for a fixed design-$E_b/N_0$. The constructed lookup tables are then stored and applied for all $E_b/N_0$. Hence, the lookup table construction has to be done only once and offline. 
	  
\section{Results and Discussion}
In this section, we present and discuss results from a bit error rate simulation of an exemplary non-binary LDPC code over Galois field GF$(4)$. The code was taken from \cite{MacKay_DB} and has length $N_\text{v}=816$, code rate $R_\text{c}=0.5$, variable node degree $d_\text{v}=3$ and check node degree $d_\text{c}=6$.

\begin{table*}[t]
\caption{Simulation parameters}
\centering
\begin{tabular}{lcccc}
decoder&check node operation&variable node operation& bits/element exchanged messages &check node operation computational complexity\\
\hline\hline
sum-product& FHT& multiplication & 64 bit &$\mathcal{O}(d_c q (\log_2 q+d_c))$\\
\hline
log-max& $\max^*()$ \cite{Wymeersch.2004} & addition & 64 bit & $\mathcal{O}(d_c q^2)$\\
\hline
proposed& lookup table & lookup table& 3 bit & $\mathcal{O}(d_c)$\\
\end{tabular}
\label{tab:params1}       
\vskip-2pt
\end{table*}

\begin{table}
\centering
\caption{Total memory amount of lookup tables in the information bottleneck decoder per iteration. }
\begin{tabular}{lcccc}
Lookup table &cardinality&table size\\
\hline\hline
Check node \circled{1}, \circled{3}& $|\mathcal{T}_\text{mult}|=256$& $3.04$ kB\\
\hline
Check node \circled{2}& $|\mathcal{T}_\text{conv}|=512$ & $129.02$ kB\\
\hline
Variable node & $|\mathcal{T}_\text{var}|=512$ & $82.94$ kB\\
\hline\hline
Total & & $215.00$ kB
\end{tabular}
\label{tab:params2}
\vskip-10pt
\end{table}

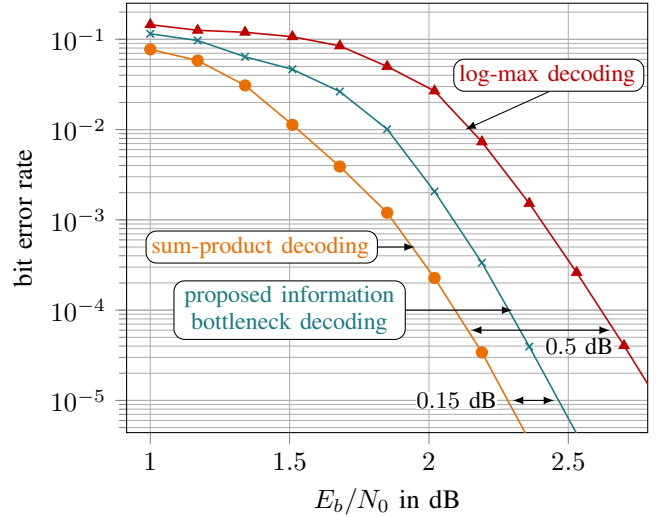
\begin{figure}
    \centering
\begin{tikzpicture}

\begin{axis}[
xmin=0.915, xmax=2.785,
ymin=4.4e-06, ymax=0.250798370513064,
ymode=log,
tick align=outside,
tick pos=left,
xmajorgrids,
xminorgrids,
x grid style={white!69.01960784313725!black},
ymajorgrids,
yminorgrids,
y grid style={white!69.01960784313725!black},
legend style={at={(0.03,0.03)}, anchor=south west, draw=white!80.0!black},
legend cell align={left},
ylabel= bit error rate,
xlabel=$E_b/N_0$ in dB,
]
\addplot [semithick, mark=otimes*, DarkOrange, solid]
table {%
1 0.077359068627451
1.17 0.0581680346557228
1.34 0.0308490680222706
1.51 0.0112900363765416
1.68 0.00389295503147496
1.85 0.00119915444687904
2.02 0.000227191333154266
2.19 3.40341515763012e-05
2.36 3.60575767385361e-06
2.53 -1
2.7 -1
};
\addplot [semithick,mark=triangle*, DarkRed, solid]
table {%
1 0.145328719723183
1.17 0.125735294117647
1.34 0.119631185807656
1.51 0.106846932321316
1.68 0.0845179738562092
1.85 0.0500288350634371
2.02 0.0267763019186169
2.19 0.00731360138992306
2.36 0.00152034598502453
2.53 0.000260755963583136
2.7 4.04583964621007e-05
2.9 4.04583964621007e-06
};
\addplot [semithick, DarkGreen, mark=x]
table {%
1 0.114973262032086
1.17 0.0969551282051282
1.34 0.0640082956259427
1.51 0.046499260081391
1.68 0.0263667440660475
1.85 0.0101052716168435
2.02 0.0020620338815627
2.19 0.000335118409615682
2.36 3.95313148651737e-05
2.53 4.28866285913352e-06
2.7 -1
};

\coordinate (spypoint) at (axis cs:4.13,3e-4);
\coordinate (magnifyglass) at (axis cs:4.7,8e-2);

\tikzstyle{mybox4} = [draw=black, fill=white, thin,
rectangle,rounded corners, inner sep=2pt, inner ysep=2pt]
\node [mybox4](boxlog) at (axis cs:2.43,4e-2){%
	\begin{minipage}{0.13\textwidth}
	\begin{flushleft}
	\small {\color{DarkRed} log-max decoding}\\
	\end{flushleft} 
	\end{minipage}
};  
\draw (boxlog.south) edge[-latex] (axis cs:2.13, 1e-2);

\tikzstyle{mybox2} = [draw=black, fill=white, thin,
rectangle,rounded corners, inner sep=2pt, inner ysep=2pt]
\node [mybox2](boxbp) at (axis cs:1.4,5e-4){%
	\begin{minipage}{0.16\textwidth}
	\begin{flushleft}
	\small {\color{DarkOrange}sum-product decoding}\\
	\end{flushleft}
	\end{minipage}
};  

\draw (boxbp.east) edge[-latex] (axis cs:1.95,5e-4);

\tikzstyle{mybox3} = [draw=black, fill=white, thin,
rectangle,rounded corners, inner sep=2pt, inner ysep=2pt]
\node [mybox3](boxib) at (axis cs:1.5,1e-4){%
	\begin{minipage}{0.16\textwidth}
	\begin{centering}
	\small {\color{DarkGreen}proposed information bottleneck decoding}\\
	\end{centering}
	\end{minipage}
};  

\draw (boxib.east) edge[-latex] (axis cs:2.3,1e-4);

\node (GI5a) [fill=white, text=black, inner sep = 0pt] at (axis cs:2.1,1e-5) {\small $0.15$ dB};   
\draw (axis cs:2.3,1e-5) edge[latex-latex] (axis cs:2.45,1e-5);

\node (GI5b) [fill=white, text=black, inner sep = 0pt] at (axis cs:2.54,4e-5) {\small $0.5$ dB};   
\draw (axis cs:2.15,6e-5) edge[latex-latex] (axis cs:2.65,6e-5);

\end{axis}


%

\end{tikzpicture}
    \vskip -5pt
    \caption{Bit error rate performance of our proposed decoder and reference systems with properties summarized in Table \ref{tab:params1} and $i_\text{max}=40$.} 
    \label{fig:BER}
    \vskip -10pt
\end{figure}

The obtained bit error rates for sum-product decoding using FHT \cite{Declercq.2007}, log-max decoding \cite{Wymeersch.2004} and the proposed information bottleneck based decoding are depicted in Figure \ref{fig:BER}.
The channel quantizer described in Section \ref{sec:qtzr} was used with output cardinality $|\mathcal{T}_\text{chan}|=128$ corresponding to 7 bit quantization. 
For the sum-product and the log-max decoder, the symbol probabilities $p(c_k|t_k)$ were used for decoding. In contrast, 
the information bottleneck decoder worked directly on the quantization indices $t_k$.

All decoders performed a maximum of $i_\text{max}=40$ iterations. 
The information bottleneck decoder was constructed for a design-$E_b/N_0$ of $1.5$ dB.
The most important parameters of the applied decoders are summarized in Table \ref{tab:params1} for a quick overview.

In the information bottleneck decoder, only integer-valued indices from the sets $\mathcal{T}_\text{mult}$ and $\mathcal{T}_\text{var}$ are used as messages instead of probability vectors.
In literature, the precision of the exchanged probability vectors is often provided in bits \textit{per field element} \cite{Wymeersch.2004b}. Thus, for a fair comparison, the cardinalities summarized in Table \ref{tab:params2}, correspond to a maximum of 3 bits per field element (cf. Table \ref{tab:params1} column 4). 


In Figure \ref{fig:BER} the sum-product algorithm serves as a benchmark with the best bit error rate performance, but at the same time, it has the highest computational complexity (cf. Table \ref{tab:params1}). Although all applied operations
in the information bottleneck decoder are simple lookups, the decoder performs only $0.15$ dB worse than the benchmark. Despite the fact that the log-max decoder uses conventional arithmetic and double-precision precision message representation, it is clearly outperformed by the proposed information bottleneck decoder. 
We emphasize that for the non-binary case the applied lookup tables completely replace all arithmetical operations such as convolution of probability vectors and multiplication. The processing of probability vectors simplifies to lookups of scalar integers in pre-generated tables. 
During our work, we noticed that the cardinalities in the information bottleneck algorithms required to obtain performance close to sum-product decoding grow with increasing Galois field order. For the considered GF(4) LDPC code, the amount of memory that is required to store the lookup tables is provided in Table \ref{tab:params2}. It can be seen that for the considered decoder, $215.00$ kilobytes (kB) are needed per iteration. 
This amount of memory required can be justified by the huge savings in computational complexity.




\label{sec:results}


\section{Conclusion}	
In this paper, we leveraged the information bottleneck's ability to preserve relevant information to overcome the main computational burdens of non-binary LDPC decoding. Motivated by the preliminary results for binary LDPC codes, 
we have presented a complete framework to design check node and variable node operations which replace all arithmetic operations using only lookups in non-binary LDPC decoders. We provided a step-by-step conversion of the conventional sum-product algorithm resulting in an information bottleneck decoder which performs only $0.15$ dB worse than the sum-product algorithm and outperforms the log-max algorithm. In addition, a discrete density evolution scheme for the proposed 
decoding method was sketched.
Future work should investigate possibilities of efficient lookup table implementation for the proposed decoding scheme. In our opinion, the combination of promising bit error rate curves and simple lookup operations already motivates further work on the information bottleneck principle in decoding of non-binary LDPC codes. Such a simple scheme for decoding could drastically increase applicability of non-binary LDPC codes in many practical scenarios.
\label{sec:conclusion}
\bibliographystyle{IEEEtran}
\bibliography{mybib}

\end{document}